\documentclass[a4paper]{jpconf}
\usepackage{amsmath,amssymb,epsfig,bm,mathrsfs,color}
\usepackage{slashed}
\newcommand{\be}{\begin{equation}}
\newcommand{\ee}{\end{equation}}
\newcommand{\bea}{\begin{eqnarray}}
\newcommand{\eea}{\end{eqnarray}}

\newcommand{\aap}{Astron.\& Astrophys.}
\newcommand{\apj}{Astrophys. J.}

\newcommand{\prc}{Phys. Rev. C}
\newcommand{\prd}{Phys. Rev. D}

\newcommand{\vecnu}{{\bm \nu}}
\newcommand{\vecv}{{\bm v}}
\newcommand{\vecD}{{\bm D}}
\newcommand{\vecF}{{\bm F}}
\newcommand{\vecA}{{\bm A}}
\newcommand{\vecnabla}{{\bm \nabla}}
\newcommand{\vecp}{{\bm p}}

\definecolor{red}{rgb}{0.8,0,0}
\definecolor{orange}{rgb}{0.8,0.2,0.0}
\definecolor{blue}{rgb}{0.3,0.0,0.8}

\def\apj{ApJ~}%
\def\apss{Ap\&SS}%
\def\aap{A\&A~ }%
\def\mnras{MNRAS~}%
\def\prc{Phys.~Rev.~C~}%
\def\prd{Phys.~Rev.~D~}%
\def\ssr{Space~Sci.~Rev.}%
\begin{document}
\title{From microphysics to dynamics of magnetars}

\author{Armen Sedrakian}
\address{Frankfurt Institute for Advanced Studies, 
D-60438 Frankfurt-Main, Germany,\\ Institute for Theoretical Physics, J.~W.~Goethe-University, D-60438 Frankfurt-Main, Germany}
\ead{sedrakian@fias.uni-frankfurt.de}

\author{Xu-Guang Huang}
\address{Physics Department and Center for Particle Physics and Field Theory, Fudan University, Shanghai 200433, China}
\ead{huangxuguang@fudan.edu.cn}

\author{Monika Sinha}
\address{Indian Institute of Technology Rajasthan, Old Residency Road, Ratanada, Jodhpur 342011, India}
\ead{monikasinha@gmail.com}

\author{John W. Clark}
\address{Department of Physics and McDonnell Center for the Space
Sciences, Washington University, St.~Louis, Missouri 63130, USA\\
Centro de Ci\^encias Matem\'aticas,
University of Madeira, 9000-390 Funchal, Portugal
}
\ead{jwc@wuphys.wustl.edu}

\begin{abstract}
  MeV-scale magnetic fields in the interiors of magnetars suppress the
  pairing of neutrons and protons in the $S$-wave state. In the case
  of a neutron condensate the suppression is the consequence of the
  Pauli-paramagnetism of the neutron gas, i.e., the alignment of the
  neutron spins along the magnetic field. The proton $S$-wave pairing
  is suppressed because of the Landau diamagnetic currents of protons
  induced by the field.  The Ginzburg-Landau and BCS theories of the
  critical magnetic fields for unpairing are reviewed. The
  macrophysical implications of the suppression (unpairing) of the
  condensates are discussed for the rotational crust-core coupling in
  magnetars and the neutrino-dominated cooling era of their thermal
  evolution.
\end{abstract}

\section{Introduction}\label{sec:1}
\vskip 0.3cm The powerful X-ray and soft $\gamma$-ray outburst activity 
observed in a number of astrophysical point sources has been attributed 
to the magnetic energy stored in compact objects. The {\it magnetar} 
interpretation of these observational phenomena implies magnetic fields 
larger by a factor $10^3$ than the fields deduced for rotationally 
powered pulsars $B\sim 10^{12}$ G~\cite{1995MNRAS.275..255T}.  The
interior fields of magnetars cannot be measured, but it has been
frequently conjectured that they are stronger than their surface 
fields by several orders of magnitude; for recent reviews
see~\cite{2015RPPh...78k6901T,2015SSRv..191..315M}. Large interior
fields affect the equation of state of nucleonic and quark matter, 
potentially endangering the hydrostatic equilibrium of compact stars. 
In fact, observationally significant modification of gross parameters of 
compact stars (mass, radius, moment of inertia, etc.) requires extremely 
high fields $B\sim 10^{18}-10^{19}$~G, for which the stars are close to loss 
of hydrostatic equilibrium as derived from the virial theorem
\cite{1953ApJ...118..116C,1991ApJ...383..745L} and general relativistic 
numerical models 
\cite{1995A&A...301..757B,2012MNRAS.427.3406F,2015MNRAS.447.3785C}.

Those physical phenomena in dense and strongly interacting matter
which are controlled by the form of the quasiparticle spectrum in the
vicinity of the Fermi surface are affected by much lower fields of
the order $B\sim 10^{16}-10^{17}$~G. For such fields, electromagnetic
interactions become of the order of the nuclear scales $\sim$ MeV
characterizing low-energy fermionic excitation spectrum of nucleons 
(as opposed to the high-energy scale set by the Fermi energy). An 
example to be addressed below is pairing in nucleonic condensates, 
which is characterized by the MeV scale. Likewise, neutrino transport 
and radiation are dominated by this scale throughout the early evolution 
of compact stars as they cool via neutrino radiation from their interiors. 
Accordingly, we will also discuss the effects of magnetic fields on the 
neutrino radiation processes. 

It is useful at this point to make the notion of MeV-scale magnetic 
fields more precise.  The interaction energy of the magnetic field 
with the nucleon spin is $\mu_NB$, where $\mu_N=e\hbar/2m_p$ 
is the nuclear magneton. Substituting the values of constants we find 
that $\mu_NB\simeq \pi (B/10^{18}~ \textrm{Gauss})$~MeV, i.e., fields 
of the order of $10^{17}$~G would substantially affect the MeV-scale
pairing via the spin--$B$-field interaction.  In the case of charged
particles, the relevant magnetic-field scale follows from the Landau
criterion for the critical velocity $v_s\sim \Delta/p_\perp$, where
$\Delta$ is the pairing gap and $p_\perp$ is the characteristic 
momentum in the plain orthogonal to the field, restricted to
$p_\perp \le p_F$ by the Fermi momentum $p_F$.  The corresponding 
energy scale is then given by $v_sp_F \sim \pi (v_s/c) 
(\xi/10~\textrm{fm}) (B/10^{16}~\textrm{Gauss})$ MeV, assuming 
that the Larmor radius $p_F c/eB$ is of the order of the coherence 
length $\xi\simeq 10$~fm and $v_s/c\le 0.3$. Thus, the range of 
characteristic fields relevant to the MeV-scale physics in the
vicinity of the Fermi surface of nucleons is $10^{16}\le B\le 10^{17}$~G.  
As will be shown, this is indeed confirmed by explicit calculations. 

The purpose of this review is to give a concise account of the
progress achieved in understanding the effect of MeV-scale
magnetic fields on nucleonic
pairing~\cite{2014arXiv1403.2829S,2015PhRvC..91c5805S,2016PhRvC..93a5802S}
and its consequences for neutrino radiation~\cite{2015PhRvC..91c5805S}
and crust-core coupling in magnetars~\cite{2016A&A...587L...2S}. The
broad and actively pursued subject of the equation of state of various
phases of dense matter in much stronger magnetic fields (comparable to
the Fermi energy scale of fundamental fermions) is not discussed here.

This review is organized as follows. Section~\ref{sec:Protons}
summarizes the Ginzburg-Landau theory of charged (proton) and neutral
(neutron) superfluid mixtures in the cores of magnetars and addresses
the critical magnetic field for destruction of proton superconductivity. 
Section~\ref{sec:Neutrons} discusses the suppression of pairing in 
the neutron $S$-wave condensate (relevant for crusts of magnetars). 
Section~\ref{sec:Dynamics} focuses on implications the suppression of 
the proton and neutron pairing for dynamical coupling between the 
core and the crust of a magnetar. Finally, Sec.~\ref{sec:Neutrinos} 
considers the modifications of the neutrino emission induced by the 
magnetic fields.  Section~\ref{sec:Remarks} contains some concluding 
remarks.

\section{Unpairing of nucleonic condensates in magnetars}
\label{sec:Unpairing}
\vskip 0.3cm 

MeV-scale magnetic fields can destroy the superconducting coherence
that is required for the formation of condensates in nucleonic
matter. The interaction of the magnetic field with the neutron or 
proton spin induces an imbalance in the number of spin-up and
spin-down particles, which implies that the Cooper pairing will be
suppressed, because not all spin-up particles will find spin-down
``partners''~\cite{2016PhRvC..93a5802S}. This so-called {\it Pauli
paramagnetic suppression} acts for both proton and neutron
condensates, but is the dominant suppression mechanism only in the
case of neutrons. The proton condensate is already suppressed by
a smaller magnetic field due to a different mechanism, associated
with the Larmor motion of protons in the magnetic field, i.e.,
originating from the interaction of the charge of the proton with the
$B$-field~\cite{2014arXiv1403.2829S,2015PhRvC..91c5805S}. The
following Sections review these mechanisms in turn.

\subsection{Critical unpairing of proton condensate}
\label{sec:Protons}
\vskip 0.3cm 

The BCS superconductors are characterized at least by three distinct
length scales: (i) the {\it London penetration depth} $\lambda$, (ii) 
the {\it coherence length} $\xi$, and (iii) the interparticle distance 
$d$. It will be assumed henceforth that the last scale is much smaller 
than the other two in the problem, i.e., the superconductor is in the weakly
coupled regime. The ratio of the remaining two scales defines the type
of the superconductivity via the Ginzburg-Landau (GL) parameter,
$\kappa= \lambda/\xi$ (see, e.g., \cite{TinkhamBook}). In the range
$ 1/\sqrt{2} < \kappa <\infty,$ the material is a type-II
superconductor; otherwise it is type-I.  In
type-II superconductors the magnetic field is carried by electromagnetic
vortices with quantum flux $\phi_0 = \pi/e$ (here and below
$\hbar=c=1$), while the field forms domain structures in a type-I
superconductor.  These two scales also define three distinct 
magnetic-field scales when combined with the flux quantum:
$$ 
H_{c1} \simeq {\phi_0}{\lambda^{-2}},\quad 
H_{cm}\simeq{\phi_0}({\xi\lambda})^{-1}, \quad 
H_{c2}\simeq {\phi_0}{\xi^{-2}}.
$$
In type-II superconductors the hierarchy of these fields is 
$H_{c1}\le H_{cm}\le H_{c2}$ when $\kappa \ge 1$.  At and above 
$H_{c1}$ the creation of a single flux-tube (Abrikosov quantum vortex) 
is energetically favorable. The field $H_{cm}$ is the thermodynamical 
magnetic field whose energy density is equal to the difference in 
the energy densities of superconducting and normal states. 
Finally, $H_{c2}$ corresponds to the field at which superconductivity 
disappears; physically the density of flux-tubes in such a 
magnetic field is so high that the normal vortex cores overlap.

The GL theory of neutron-proton superfluid mixtures is based on the
functional~\cite{1980Ap.....16..417S,1984ApJ...282..533A,2005PhRvC..72e5801A}
\bea
\label{GL_Functional}
\mathscr{F}
[\phi,\psi] = \mathscr{F}_n[\phi,\psi]  + \mathscr{F}_{p}[\phi,\psi] 
+\frac1{4m_p}\vert \vecD\psi\vert^2 + \frac{B^2}{8\pi},
\eea
where $\psi$ and $\phi$ are the condensate wave functions for protons
and neutrons, $\vecD = -i\vecnabla - 2e \vecA$ is the gauge-invariant
derivative, $m_p$ is the proton mass and indices $p$ and $n$ label the
quantities referring to the neutron and proton condensates. If we are interested 
only in the $H_{c2}$ field, the wave function is small and the 
functional for the proton condensate can be written at temperature $T$ 
and critical temperature $T_{cp}$ of the superconducting phase 
transition as
\bea
\mathscr{F}_{p}[\phi,\psi] = \alpha\tau \vert \psi\vert^2 +\frac{b}{2}\vert
\psi\vert^4
+b'\vert \psi\vert^2 \vert \phi\vert^2,
\eea
where $\tau = (T-T_{cp})/T_{cp}$.  The quantities $\alpha$ and $b$ are the
coefficients of the GL expansion, while $b'$ describes the density-density
coupling between the neutron and proton condensates. The current-current 
coupling, i.e., the {\it entrainment} of the proton condensate by the 
neutron condensate, can be absorbed in the effective gauge potential 
$\vecA$~\cite{2015PhRvC..91c5805S,1980Ap.....16..417S}.

The variations $\delta \mathscr{F}[\phi,\psi] /\delta \psi =0$,
$\delta \mathscr{F}[\phi,\psi] /\delta \phi =0$, and
$\delta \mathscr{F}[\phi,\psi] /\delta \vecA =0$ constitute the
coupled equations of motion describing the mixture of condensates.
Assuming that the neutron condensate is static and homogeneous
and that $\vecA$ is locally linear in the coordinates (constant
$B$-field), the linearized GL equations provide the value 
\bea
\label{eq:Hc2_1}
H_{c2} &=&\frac{\phi_0}{2\pi\xi_p^2} \left[1 + \beta(b')\right],
\eea 
of the critical field $H_{c2}$~\cite{2015PhRvC..91c5805S},
where $m_p\vert\alpha\tau\vert = (2\xi_p)^{-2}$. The coupling of the
proton condensate to the neutron condensate enhances the critical
field by a factor $\beta\simeq 0.2$. 
\begin{figure}[t]
\centering 
 \includegraphics[width=8cm]{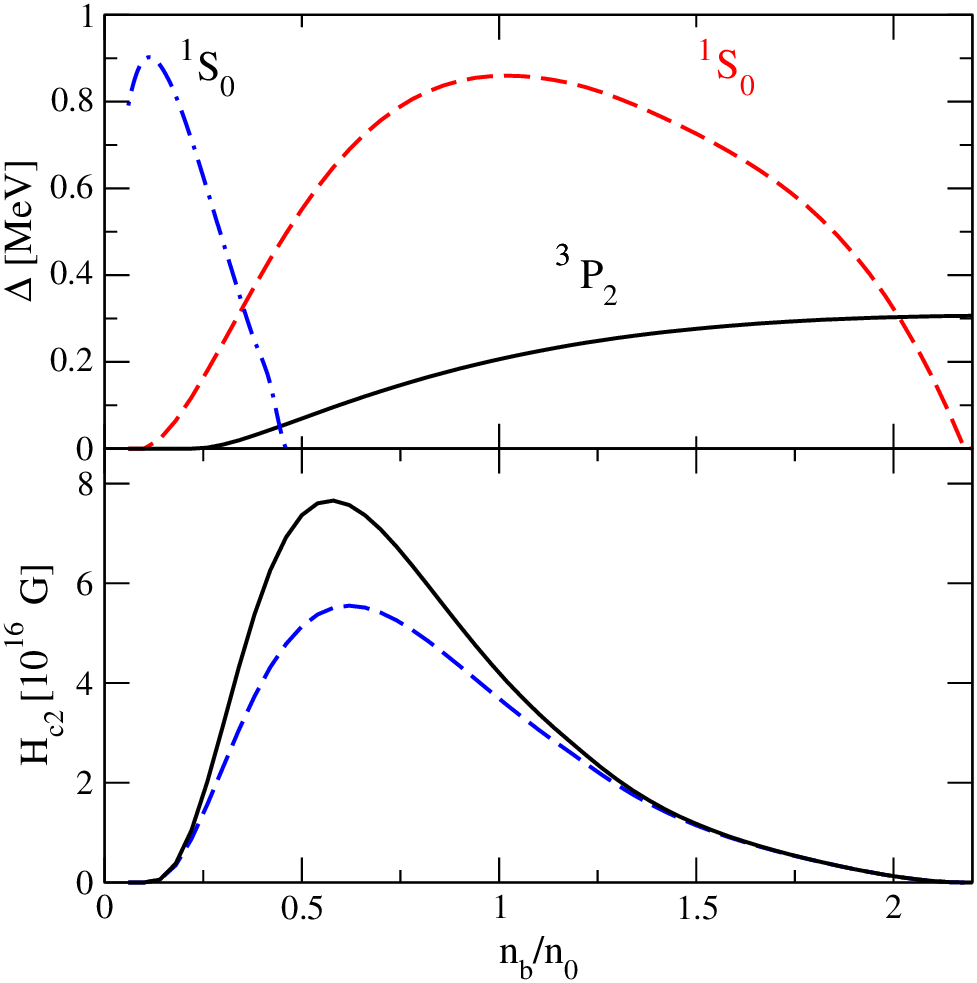}
 \caption{Conjectured layered structure of a magnetar with a constant
   magnetic field in the core (indicated by the dashed line). To the right
   of the intersection of this field with $H_{c2}$, the proton fluid is
   non-superconducting; to the left, i.e., at $B\le H_{c2}$, it is
   superconducting. The crust contains a homogeneous $B$-field. The
   density is given in units of the nuclear saturation density $n_0$.}
 \label{fig:Hc2_protons}
\end{figure}
The dependence of the $H_{c2}$ field on density is illustrated in
Fig.~\ref{fig:Hc2_protons} according to~\cite{2015PhRvC..91c5805S}.
The following features are notable. The maximum of $H_{c2}$ is
attained close to the crust-core interface at a density
$n_{\rm b} = 0.5 n_0$, where $n_0$ is the nuclear saturation density.
Fig.~\ref{fig:Hc2_protons} shows the conjectured layered structure of
a magnetar: (a) an inner core void of superconductivity, (b) an outer
core threaded by flux tubes, and (c) a crust containing a homogeneous
field of magnitude $B$.  If the field $B\ge {\rm max}\, [H_{c2}]$ the
intermediate flux-carrying region disappears, i.e., superconductivity
in a magnetar is completely destroyed. The critical fields in a
similar formalism for superfluid-superconducting mixture were
studied recently in \cite{2016arXiv161201865H}.

\subsection{Critical unpairing of neutron $S$-wave condensate
}\label{sec:Neutrons}
\vskip 0.3cm 

The nature of the suppression of pairing in the neutron condensate differs
from that in the proton condensate, because charge-neutral neutrons
interact with the $B$-field via their spin magnetic moment. This has a
destructive effect on $S$-wave neutron Cooper pairs, which involve
spin-up and down partners. Clearly, a large enough magnetic field will
quench pairing completely.  In the following, this field will be referred
to as $H_{c2}$, as no confusion should arise with the analogous quantity for
protons.

$S$-wave pairing is relevant for the crusts of magnetars, i.e.,
the low-density regime below the saturation density of symmetrical
nuclear matter.  At higher densities the dominant pairing state in
neutron matter shifts to the $^3P_2$-$^3F_2$ channel, which pairs
neutrons in a total spin-1 state (see~\cite{2003NuPhA.720...20Z,2016JPhCS.702a2012C}
for a review). In this case, the spin-polarizing effect of the magnetic 
field on the internal structure of the spin-1 pairs is 
nondestructive~\cite{1980PhRvD..21.1494M,1982PhRvD..25..967S,2014EL....10552001T}.

\begin{figure}[t]
\centering
 \includegraphics[width=10cm]{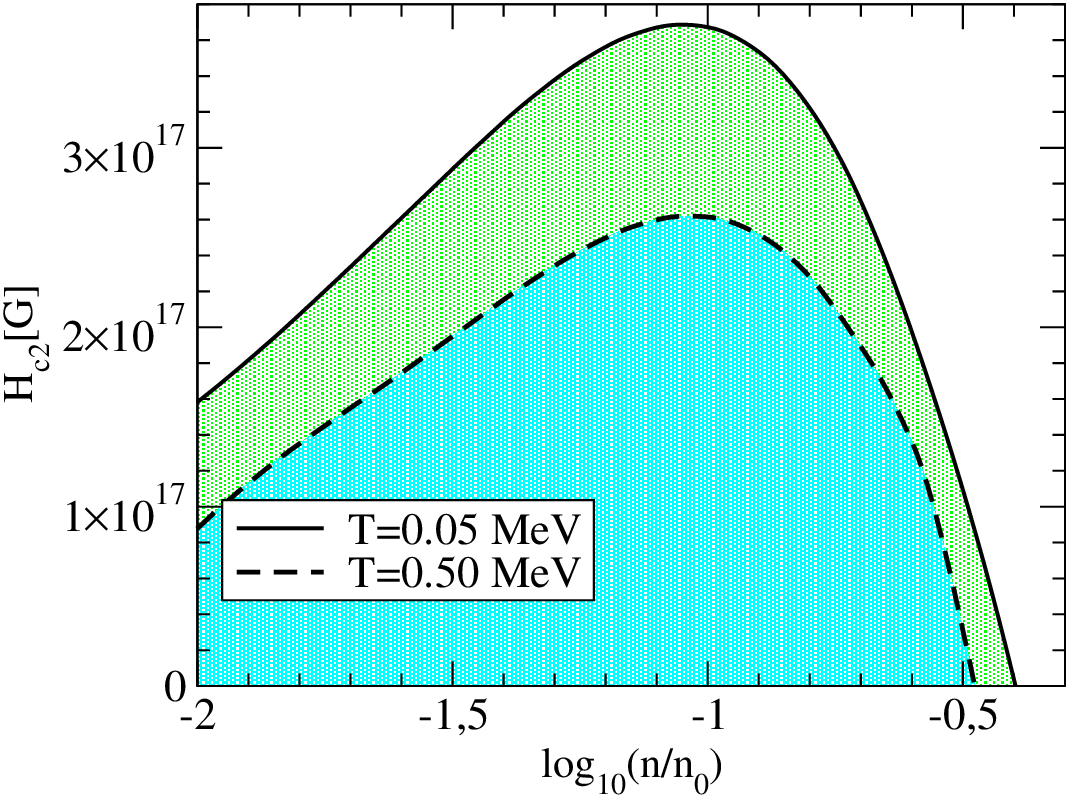}
 \caption{Critical unpairing magnetic field $H_{c2}$ for the neutron
   condensate due to spin alignment in this field, for
   two temperatures indicated in the plot. The density is given in
   units of the nuclear saturation density $n_0$.  }
 \label{fig:Hc2neutrons}
\end{figure}

In Fig.~\ref{fig:Hc2neutrons} we plot values of $H_{c2}$ for the neutron 
condensate as a function of density, determined based on a 
phase-shift-equivalent nucleon-nucleon interaction and numerical 
solutions of the BCS equations in the case of spin-polarized neutron
matter~\cite{2016PhRvC..93a5802S}. The shape of the curve reflects 
the corresponding density dependence of the pairing gap, and its
temperature dependence follows the BCS prediction: it is largest at
$T =0$ and decreases as the pairing gap decreases with increasing
temperature.
Thus, if the local field in a magnetar crust exceeds the value $H_{c2}$, 
the magnetic field will destroy the condensate.  In the specific model 
of \cite{2016PhRvC..93a5802S}, the neutron fluid in magnetars will be 
non-superfluid (i.e., in a normal phase) for $B> 2.6\times 10^{17}$ G. 

The non-superfluidity or partial superfluidity of magnetars will
clearly have important implications for an array of microphysical
quantities of neutron star crusts. These include their neutrino
emissivities and transport properties, and consequently the thermal
relaxation and dynamical coupling time scales that are important, 
most notably, for the damping of stellar oscillations and the
interpretation of rotational anomalies such as glitches and
anti-glitches. Finally, it is to be noted that the Pauli paramagnetic
destruction mechanism discussed here for $S$-wave paired neutrons will
apply as well to $S$-wave paired protons; however, the diamagnetic
mechanism mentioned in the preceding section is more important for
protons.

\section{Crust-core coupling time scales in
  magnetars}\label{sec:Dynamics}
\vskip 0.3cm 

We turn now to implications of the unpairing effect for the rotational
coupling of a neutron superfluid in magnetar cores. In
\cite{2016A&A...587L...2S} it was argued that unpairing offers a new
channel for coupling of the electron-proton plasma to the neutron
vorticity in the core: if protons are unpaired then they are available
for scattering off neutron quasiparticles in the vortex cores. This
process is much more effective than the scattering of electrons off
magnetized neutron vortices by electromagnetic forces, which is the
dominant process in the type-I superconducting case. In the more
realistic case of type-II superconductivity, the coupling mechanism
can be more complicated, because of the non-negligible interactions
between the protonic flux-tubes and the neutron vortices.

The time scale of dynamical coupling of the superfluid to the normal
plasma is important for the interpretation of rotational
irregularities of magnetars, including glitches, anti-glitches,
post-glitch relaxation, and non-axisymmetric motions such as
precession. The influence of an interior fluid on precession has
been discussed extensively in the literature for the case of ordinary 
neutron stars (see \cite{2007Ap&SS.308..435L} and references therein). 
In the case of magnetars, the coupling time scale of the $P$-wave 
neutron superfluid when the protons form a normal fluid is an important 
ingredient of such considerations~\cite{2016A&A...587L...2S}.

To illustrate this point, assume a constant field in the core of a
magnetar and a fully unpaired proton fluid.  The field will then couple to 
the electron fluid on plasma time scales, which are much shorter than the
hydrodynamical time scales. Therefore, the unpaired core of a magnetar
can be considered as a two-fluid system consisting of a superfluid neutron
condensate component and a normal component formed by the proton and 
electron fluids. 

The neutron superfluid rotates by forming an array of quantized
vortices with areal number density 
\bea
\label{eq:vortex_number}
N_n = \frac{2\Omega}{\omega_0}, \quad \omega_0 =  \frac{\pi}{m_n},
\eea
where $m_n$ is the bare neutron mass, $\Omega$ the rotation
frequency of the star, and $\omega_0$ the quantum of neutron
circulation.  As seen from Eq.~\eqref{eq:vortex_number}, any changes 
in the rotation frequency of a magnetar must be accompanied by changes 
in the number of neutron vortices. Because the vortices are created and
destroyed at the interfaces, they need to move in the bulk of the
superfluid to respond to variations in $\Omega$. The velocity of a
vortex $\vecv_L$ is determined by the equation of motion 
\be 
\label{eq:force_balance} 
\rho_n\omega_0 [(\vecv_S-\vecv_L)\times
\vecnu] - \eta (\vecv_L-\vecv_N) =0.  
\ee 
This equation reflects the balance of forces acting on a vortex segment;
the first term of the right is the Magnus force, and the second is 
the frictional force between the vortices and the normal liquid,
with $\rho_n$ denoting the mass density of the superfluid component, 
$\vecv_N$ is the velocity of the normal component and $\eta$ is 
the coordinate-dependent longitudinal friction coefficient and
$\vecnu$ is the unit vector along the vortex circulation. The 
frictional force due to scattering of normal quasiparticles 
(electrons, muons, and unpaired protons) is given by 
\be \label{eq:friction_force}
\vecF = \frac{2}{\tau N_n} \int f (\vecp, \vecv_L) \vecp
\frac{d^3p}{(2\pi\hbar)^3} = -\eta \vecv_L,
\ee
where $f(\vecp, \vecv_L)$ is the non-equilibrium distribution
function in the frame where $\vecv_N=0$.  Assuming a small perturbation,
this function can be expanded about the equilibrium distribution Fermi 
function $f_0$ to obtain
$f(\vecp, \vecv_L) 
= f_0(\vecp)+ (\partial f_0/ \partial \epsilon)(\vecp\cdot \vecv_L).$
%
For strongly degenerate systems, $\partial f_0/\partial \epsilon\simeq 
-\delta(\epsilon-\epsilon_{Fp})$ and 
\begin{figure}[t]
\begin{center}
\vskip 1.cm 
\includegraphics[width=9cm,height=7cm]{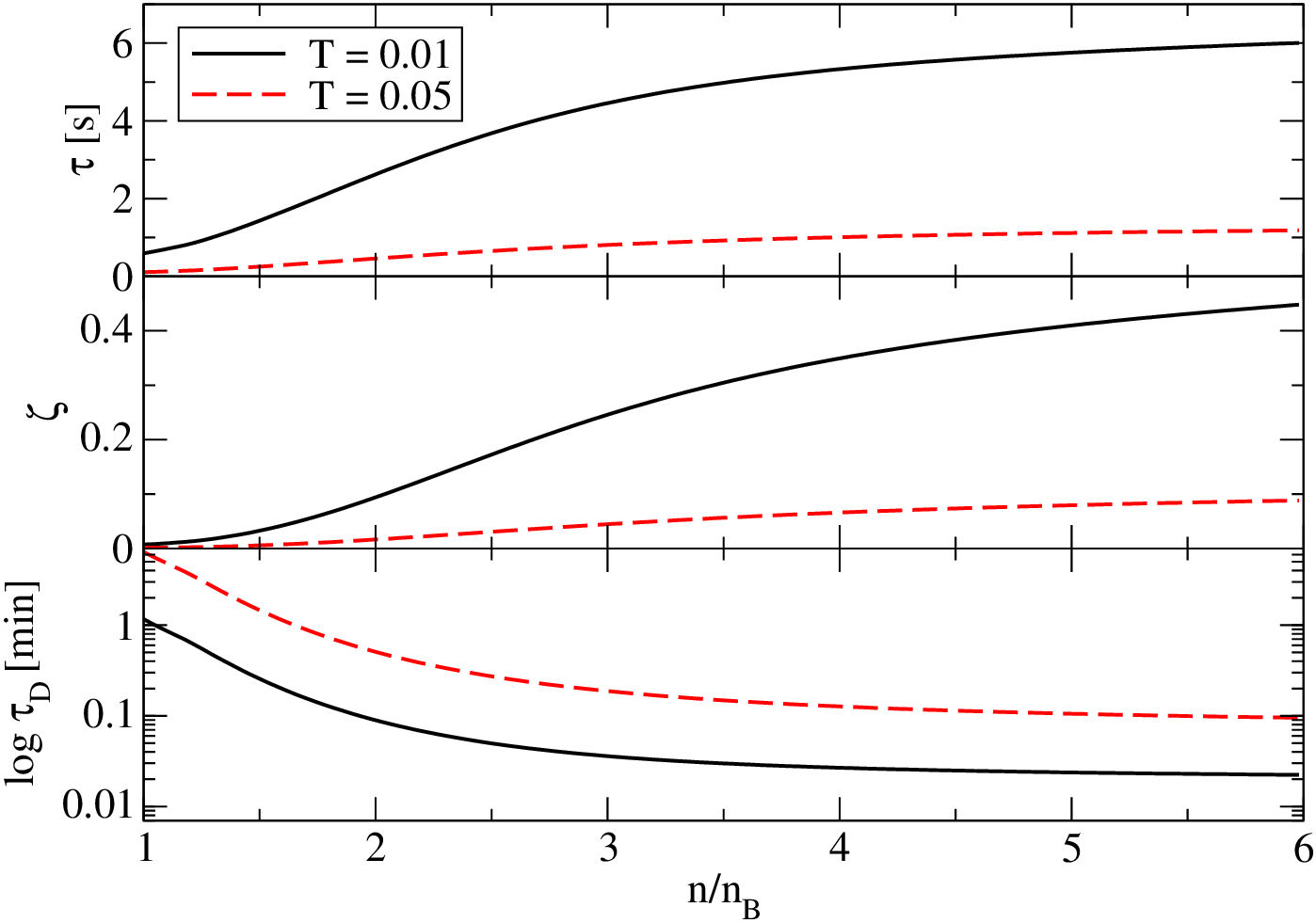}
\caption{Relaxation time, drag-to-lift ratio, and 
  dynamical coupling time scale \eqref{eq:taudynamical} as functions
  of baryon density in the stellar core for temperatures $T=0.01$ and 
  $0.05$ MeV (or equivalently $T= 1.2\times 10^{8}$ K and
  $T= 5.8\times 10^{8}$ K).  The computations are carried out with an
  angle-independent neutron-proton cross section $\sigma\simeq 60$
  fm$^2$ and for a rotation frequency $\Omega = 1$~Hz.  Note that these
  results apply only if the proton fluid is unpaired at a given
  density. }
\label{fig:tauD}
\end{center}
\end{figure}
$ \eta ={m_p^* n_p}/{\tau  N_n},$
where $n_p$ and $m_p^*$ are respectively the unpaired quasiparticle 
number density and effective mass.  In our example, the dominant 
process determining the relaxation time $\tau$ is the scattering 
of unpaired protons off neutron vortex-core quasiparticles; its 
rate scales with the temperature as
$ \tau^{-1} \propto T \exp[-{{\varepsilon_{1/2}^0}/T}] $ and is
proportional to the differential nuclear scattering cross-section
$d\sigma /d\Omega$ between neutrons and protons. The exponential
factor contains the energy scale $\varepsilon_{1/2}^0 = 
\pi\Delta_n^2/(4\epsilon_{Fn})$, corresponding to the lowest energy 
state of a neutron quasiparticle confined in the vortex, with 
$\Delta_n$ denoting the neutron pairing gap and $\epsilon_{Fn}$ 
the neutron Fermi energy; (see \cite{1998PhRvD..58b1301S} for 
further details.)  The dynamical coupling time of the superfluid 
to the plasma is given by
\bea
\label{eq:taudynamical}
\tau_D = \frac{1}{2\Omega}\left( \zeta + \zeta^{-1}\right) ,
\eea 
where the dimensionless {\it \textup{drag-to-lift ratio}} $\zeta$ 
is related to the dimensioned friction $\eta$ via the relation 
$\zeta = {\eta}/{\rho_n\omega_0}$.
Figure \ref{fig:tauD} summarizes the results of computations carried
out in~\cite{2016A&A...587L...2S}.  It is seen that for typical
magnetar periods of about 10 sec, i.e., for spin rotations of about
1~Hz, the unpaired core couples to the plasma on dynamical time scales
from several minutes at the crust-core boundary to a few seconds in
the high-density core. Furthermore, the values of $\zeta$ obtained
imply that the low-density outer core ($\zeta\simeq 0.2$) does not
affect free precession, while the high-density inner core
($\zeta\simeq 0.4$) can cause significant damping of precession over a
cycle. Thus, magnetar precession cannot be definitely excluded since
the values of the drag-to-lift ratio are within the range of the
crossover from undamped to damped precession. However, in the inner
core, where $\zeta$ is large, the condensate vanishes for lower
magnetic fields. We conclude that relatively low magnetic fields are
sufficient to damp any free precession in magnetars.  Of course, this 
argument applies only to free precession.  Magnetic deformations of
magnetars can be a continuous source of excitation of
precession~\cite{2003MNRAS.341.1020W,2015MNRAS.449.2047L}. The macroscopic arrangement of
the magnetic field in the case of type-II superconducting stars and
its implications for precession have been discussed
elsewhere~\cite{2008MNRAS.383.1551A,2013MNRAS.431.2986H,2014MNRAS.437..424L}.

\section{Neutrino radiation from magnetars}\label{sec:Neutrinos}
\vskip 0.3cm 

The suppression of pairing by MeV-scale magnetic fields will also have
profound consequences on the thermal evolution of magnetars, because
the dominant processes of neutrino radiation will not be suppressed by
the Boltzmann factor containing the gap in the quasiparticle spectrum
of baryons. At the same time, the processes that are intrinsic to
condensates, such as the pair-breaking emission of
neutrino-anti-neutrino pairs, will not operate by definition.

\subsection{Direct Urca process: $n\to p + e + \bar\nu_e$}
\vskip 0.3cm 

In strong $B$-fields the phase-space of nucleons is modified and the Urca
process is allowed even below the threshold 
$x_p\simeq 11\%$, where $x_p$ is the proton fraction in nucleonic
matter~\cite{1998JHEP...09..020L,1999A&A...342..192B}.
The kinematics of the Urca process in this case is conveniently
characterized by the parameter~\cite{1999A&A...342..192B}
\be
x = \left[1 - (k_{Fe}+k_{Fp})^2/k_{Fn}^2\right] N_{Fp}^{2/3},
\label{xdef}
\ee
where $k_{Fi}$, $i=e,p,n$, are the Fermi momenta of electrons ($e$),
protons ($p$) and neutrons ($n$) and
$N_{Fp} = k_{Fp}^2/2\vert e\vert B$ is the number of Landau levels
populated by protons. Clearly, for $x>0$ the Urca process is forbidden
in the $B=0$ case, but in a strong magnetic field some phase space opens
up. Then, if the Urca process operates even at a fraction of its
strength in the kinematically permitted region, it can become an
important factor in the cooling the star's core, because other
competing processes are weaker by orders of magnitude. In the case
where $x<0$, i.e., when the Urca process is kinematically permitted,
magnetic fields induce {\it de Haas-van Alfven oscillations} in the
emissivity associated with the filling of the Landau levels.

As well known, proton and neutron pairing suppress the Urca
process once nucleons make a transition to a superconducting or
superfluid state. At asymptotically low temperatures, the emissivity
is suppressed by a factor $\exp(-\Delta/T)$ for each participating
nucleon, where $\Delta$ is the relevant pairing gap and $T$ is the
temperature~\cite{1999PhyU...42..737Y,2005PhLB..607...27S,2007PrPNP..58..168S}.
Clearly, termination of neutron and proton $S$-wave pairing by the
$B$-field will mitigate this suppression. Because the neutron pairing 
gap in the $P$-wave channel is smaller than the proton pairing 
gap in the $S$-wave channel, destruction of proton superconductivity 
by a MeV-scale magnetic field will strongly modify the Urca emissivity.
Numerical examples can be found in~\cite{2015PhRvC..91c5805S}.

\begin{figure}[t]
\centering 
 \includegraphics[width=8cm]{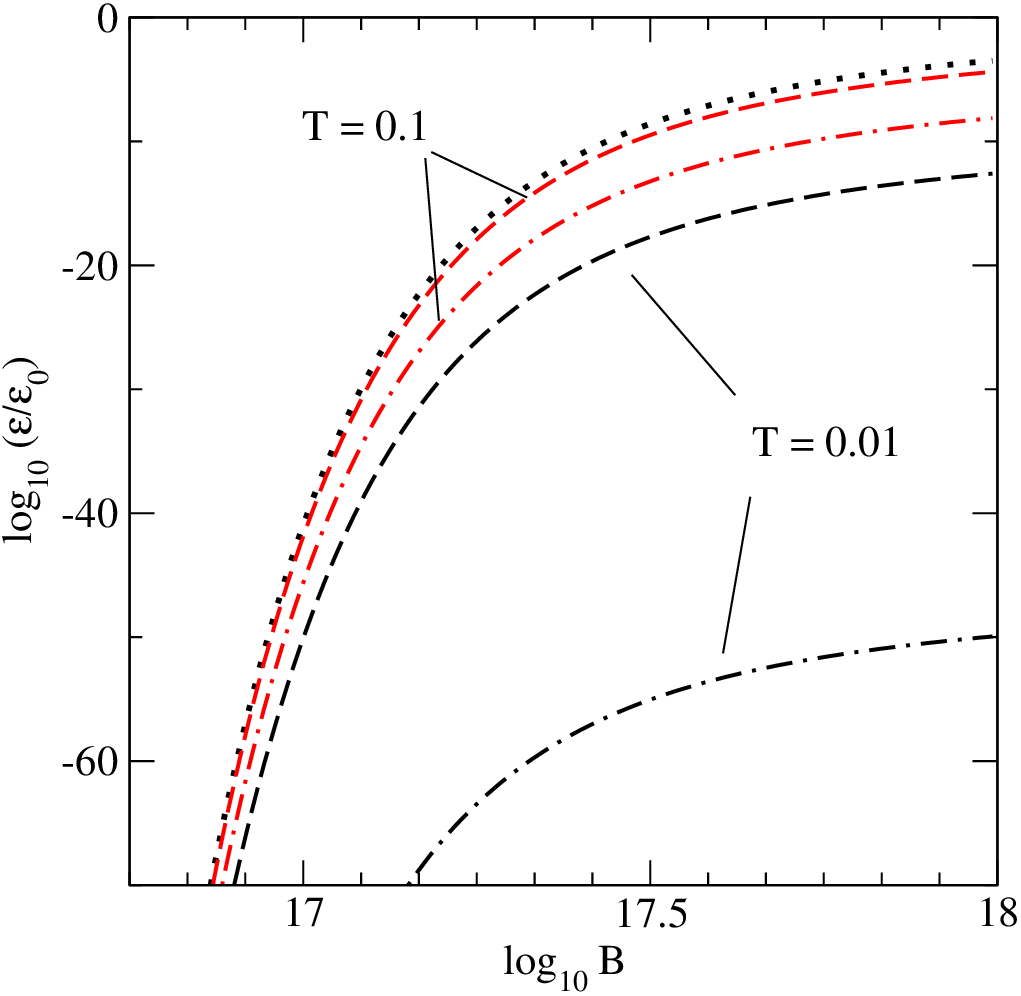}
 \caption{Magnetic-field dependence of the emissivity $\epsilon$ of the 
 Urca process (measured in units of the zero-field emissivity 
 $\epsilon_0$), as calculated at temperatures $T= 0.01$ and $0.1$ 
 MeV and density $n=n_0$ for the case $x=0$, i.e., such that emission 
 is forbidden kinematically in the zero-field limit.  Three possibilities 
 are considered: (i) normal neutron-star matter (dotted line), 
 (ii) neutrons paired but protons normal (dashed line), and (iii) both 
 neutrons and protons paired (dot-dashed lines).  The parameter $x$ of 
 Eq.~\eqref{eq:vortex_number} scales with the field $B$ according 
 to $x\propto N_{Fp}^{2/3}\propto B^{-2/3}$.}
 \label{fig:Hc2}
\end{figure}
\subsection{Field-assisted bremsstrahlung 
processes: $N \to N + \nu_f+\bar\nu_f$}
\vskip 0.3cm

The bremsstrahlung $N \to N + \nu_f+\bar\nu_f$ of neutrino pairs 
by a single nucleon (denoted $N$) is ordinarily prohibited by energy 
and momentum conservation. Accordingly, the leading charge-neutral 
mechanism for neutrino-pair production is the so-called modified 
bremsstrahlung process $N+N \to N +N+ \nu_f+\bar\nu_f$.  However, 
when the interaction energy of the $B$-field with the spin of a nucleon 
becomes of the order of the temperature, the single-nucleon bremsstrahlung 
process $N\to N+\nu+\bar\nu$ becomes kinematically possible because of 
the paramagnetic splitting of the energies of nucleons with spin-up and
spin-down in a strong $B$ field~\cite{2000A&A...360..549V}. Pairing will
suppress this process exponentially at low temperatures, as discussed
above in the case of the Urca process.  However, if unpairing by a
strong magnetic field takes place, this suppression will be nullified.

\subsection{
Pair-breaking processes: $[NN] \to N+N + \nu_f+\bar\nu_f$
}
\vskip 0.3cm

Nucleonic superfluidity results in a new class of neutrino
bremsstrahlung processes that owe their existence to the condensate.
Symbolically, these can be written as
$[NN] \to N+N + \nu_f+\bar\nu_f$, where $[NN]$ stands for a Cooper
pair. These processes are referred to as pair-breaking-formation (PBF)
processes~\cite{1976ApJ...205..541F,2006PhLB..638..114L,2007PhRvC..76e5805S,2008PhRvC..77f5808K,2009PhRvC..79a5802S,2012PhRvC..86b5803S}.
The rates of neutrino emission via the PBF processes scale as
$\epsilon \propto\Delta^2T^7$, where $\Delta$ is the pairing gap.
Therefore, the unpairing of the $S$-wave condensates will have the
plain effect of removing the PBF processes from the regions where the
field locally exceeds the unpairing fields for protons and neutrons.
Consequently, the net neutrino emission rate will be reduced
asymptotically to the value which corresponds to the PBF emission by
the $P$-wave condensate.

Assessment of the combined effect of unpairing on the cooling
of neutron stars is difficult. While it is clear how the individual
processes are affected by the strong magnetic field, their concerted
effect needs to be studied in numerical simulations.

\subsection{Specific heats}
\vskip 0.3cm

An additional ingredient that modifies the thermal evolution of
magnetars is the specific heat of the interior matter, which 
quantifies the thermal inertia of the star. As well known, in the
absence of superconductivity and superfluidity, the heat capacity 
of nucleonic fluids will scale linearly with temperature.  This 
should be contrasted with the exponential suppression of the heat 
capacity in the superconducting $S$-wave state.  In fully 
superconducting/superfluid neutron stars at low $B$-fields, 
electrons dominate the heat capacity. One can anticipate that 
for MeV-scale magnetic fields in which only protons are unpaired,
the proton and electron specific heats will decrease linearly with
temperature (as in normal Fermi liquids), whereas the heat capacity of
the superfluid neutrons will be reduced exponentially in the regions of
$S$-wave pairing and as a power-law in the regions of $P$-wave
pairing. Increase of the specific heat of the interior matter will 
act to increase the cooling time scale of a magnetar. 

In closing, it should be mentioned that other factors such as 
internal heating due to the Ohmic dissipation of the magnetic fields
in the interior of the star will be an important factor in determining
the temperature evolution of magnetars. The unpairing of protons will
imply that the resistivity due to electron-proton scattering will be
larger than in superconducting stars. Consequently, the time scales
of the Ohmic dissipation will be shorter.

\section{Final remarks}\label{sec:Remarks}
\vskip 0.3cm 

Magnetars pose new challenges at the microphysical level because the
electromagnetic interactions (e.g.\ the magnetic field-nucleon spin
coupling) become of the order of the nuclear MeV scale. As a
consequence, we find an intimate interplay between the electromagnetic
interactions and nuclear and weak processes that take place in the
vicinity of the Fermi surfaces of nucleons. This contribution has focused
on the recent progress in understanding the mechanisms of suppression
of pairing in nucleonic matter by a magnetic field and their
implications for macroscopic dynamics of magnetars, such as their
rotational and thermal evolutions.  These findings call for more
detailed studies of the {\it macroscopic dynamics} of magnetars, which
will entail the modification of the pairing properties of nucleonic
fluids.

\ack \vskip 0.3cm 

A.S.\ acknowledges the support by the DFG
(Grant No.\ SE 1836/3-2), by the Helmholtz International Center for
FAIR, and by the NewCompStar COST Action MP1304.  J.W.C.\ is grateful
for the hospitality of the Center for Mathematical Sciences of the
University of Madeira.

\section*{References}

\vskip 0.3cm
\bibliographystyle{iopart-num}

\begin{thebibliography}{10}
\expandafter\ifx\csname url\endcsname\relax
  \def\url#1{{\tt #1}}\fi
\expandafter\ifx\csname urlprefix\endcsname\relax\def\urlprefix{URL }\fi
\providecommand{\eprint}[2][]{\url{#2}}

\bibitem{1995MNRAS.275..255T}
{Thompson} C and {Duncan} R~C 1995 {\em \mnras\/} {\bf 275} 255--300

\bibitem{2015RPPh...78k6901T}
{Turolla} R, {Zane} S and {Watts} A~L 2015 {\em Reports on Progress in
  Physics\/} {\bf 78} 116901

\bibitem{2015SSRv..191..315M}
{Mereghetti} S, {Pons} J~A and {Melatos} A 2015 {\em \ssr\/} {\bf 191} 315--338

\bibitem{1953ApJ...118..116C}
{Chandrasekhar} S and {Fermi} E 1953 {\em \apj\/} {\bf 118} 116

\bibitem{1991ApJ...383..745L}
{Lai} D and {Shapiro} S~L 1991 {\em \apj\/} {\bf 383} 745--751

\bibitem{1995A&A...301..757B}
{Bocquet} M, {Bonazzola} S, {Gourgoulhon} E and {Novak} J 1995 {\em \aap\/}
  {\bf 301} 757

\bibitem{2012MNRAS.427.3406F}
{Frieben} J and {Rezzolla} L 2012 {\em \mnras\/} {\bf 427} 3406--3426

\bibitem{2015MNRAS.447.3785C}
{Chatterjee} D, {Elghozi} T, {Novak} J and {Oertel} M 2015 {\em \mnras\/} {\bf
  447} 3785--3796

\bibitem{2014arXiv1403.2829S}
{Sinha} M and {Sedrakian} A 2015 {\em Physics of Particles and Nuclei\/} {\bf
  46} 1510

\bibitem{2015PhRvC..91c5805S}
{Sinha} M and {Sedrakian} A 2015 {\em \prc\/} {\bf 91} 035805

\bibitem{2016PhRvC..93a5802S}
{Stein} M, {Sedrakian} A, {Huang} X~G and {Clark} J~W 2016 {\em \prc\/} {\bf
  93} 015802

\bibitem{2016A&A...587L...2S}
{Sedrakian} A 2016 {\em \aap\/} {\bf 587} L2

\bibitem{TinkhamBook}
{Tinkham} M 1996 {\em Introduction to superconductivity, McGraw-Hill, New
  York\/}

\bibitem{1980Ap.....16..417S}
{Sedrakyan} D~M and {Shakhabasyan} K~M 1980 {\em Astrophysics\/} {\bf 16} 417

\bibitem{1984ApJ...282..533A}
{Alpar} M~A, {Langer} S~A and {Sauls} J~A 1984 {\em \apj\/} {\bf 282} 533--541

\bibitem{2005PhRvC..72e5801A}
{Alford} M, {Good} G and {Reddy} S 2005 {\em \prc\/} {\bf 72} 055801

\bibitem{2016arXiv161201865H}
{Haber} A and {Schmitt} A 2016 {\em ArXiv e-prints\/} (\textit{Preprint}
  \eprint{1612.01865})

\bibitem{2003NuPhA.720...20Z}
{Zverev} M~V, {Clark} J~W and {Khodel} V~A 2003 {\em Nuclear Physics A\/} {\bf
  720} 20--42

\bibitem{2016JPhCS.702a2012C}
{Clark} J~W, {Sedrakian} A, {Stein} M, {Huang} X~G, {Khodel} V~A, {Shaginyan}
  V~R and {Zverev} M~V 2016 {\em Journal of Physics Conference Series\/} {\bf
  702} 012012

\bibitem{1980PhRvD..21.1494M}
{Muzikar} P, {Sauls} J~A and {Serene} J~W 1980 {\em \prd\/} {\bf 21} 1494--1502

\bibitem{1982PhRvD..25..967S}
{Sauls} J~A, {Stein} D~L and {Serene} J~W 1982 {\em \prd\/} {\bf 25} 967--975

\bibitem{2014EL....10552001T}
{Tarasov} A~N 2014 {\em EPL (Europhysics Letters)\/} {\bf 105} 52001

\bibitem{2007Ap&SS.308..435L}
{Link} B 2007 {\em \apss\/} {\bf 308} 435--441

\bibitem{1998PhRvD..58b1301S}
{Sedrakian} A 1998 {\em \prd\/} {\bf 58} 021301

\bibitem{2003MNRAS.341.1020W}
{Wasserman} I 2003 {\em \mnras\/} {\bf 341} 1020--1040

\bibitem{2015MNRAS.449.2047L}
{Lander} S~K, {Andersson} N, {Antonopoulou} D and {Watts} A~L 2015 {\em
  \mnras\/} {\bf 449} 2047--2058

\bibitem{2008MNRAS.383.1551A}
{Akg{\"u}n} T and {Wasserman} I 2008 {\em \mnras\/} {\bf 383} 1551--1580

\bibitem{2013MNRAS.431.2986H}
{Henriksson} K~T and {Wasserman} I 2013 {\em \mnras\/} {\bf 431} 2986--3002

\bibitem{2014MNRAS.437..424L}
{Lander} S~K 2014 {\em \mnras\/} {\bf 437} 424--436

\bibitem{1998JHEP...09..020L}
{Leinson} L~B and {P{\'e}rez} A 1998 {\em Journal of High Energy Physics\/}
  {\bf 9} 020

\bibitem{1999A&A...342..192B}
{Baiko} D~A and {Yakovlev} D~G 1999 {\em \aap\/} {\bf 342} 192--200

\bibitem{1999PhyU...42..737Y}
{Yakovlev} D~G, {Levenfish} K~P and {Shibanov} Y~A 1999 {\em Physics Uspekhi\/}
  {\bf 42} 737--778

\bibitem{2005PhLB..607...27S}
{Sedrakian} A 2005 {\em Physics Letters B\/} {\bf 607} 27--34

\bibitem{2007PrPNP..58..168S}
{Sedrakian} A 2007 {\em Progress in Particle and Nuclear Physics\/} {\bf 58}
  168--246

\bibitem{2000A&A...360..549V}
{van Dalen} E~N~E, {Dieperink} A~E~L, {Sedrakian} A and {Timmermans} R~G~E 2000
  {\em \aap\/} {\bf 360} 549--558

\bibitem{1976ApJ...205..541F}
{Flowers} E, {Ruderman} M and {Sutherland} P 1976 {\em \apj\/} {\bf 205}
  541--544

\bibitem{2006PhLB..638..114L}
{Leinson} L~B and {P{\'e}rez} A 2006 {\em Physics Letters B\/} {\bf 638}
  114--118

\bibitem{2007PhRvC..76e5805S}
{Sedrakian} A, {M{\"u}ther} H and {Schuck} P 2007 {\em \prc\/} {\bf 76} 055805

\bibitem{2008PhRvC..77f5808K}
{Kolomeitsev} E~E and {Voskresensky} D~N 2008 {\em \prc\/} {\bf 77} 065808

\bibitem{2009PhRvC..79a5802S}
{Steiner} A~W and {Reddy} S 2009 {\em \prc\/} {\bf 79} 015802

\bibitem{2012PhRvC..86b5803S}
{Sedrakian} A 2012 {\em \prc\/} {\bf 86} 025803

\end{thebibliography}

\providecommand{\newblock}{}

\end{document}